\documentclass[journal]{IEEEtran}

\usepackage[xindy, nonumberlist, acronym]{glossaries}
\makeglossaries

\usepackage{graphicx} %
\usepackage{hyperref}
\usepackage{url}
\usepackage[dvipsnames,table,xcdraw]{xcolor}
\usepackage[capitalize]{cleveref}

\usepackage[backend=biber,style=ieee]{biblatex}
\AtBeginBibliography{\footnotesize}

\usepackage{tikz}
\usetikzlibrary{angles}
\usetikzlibrary{arrows}
\usetikzlibrary{arrows.meta}
\usetikzlibrary{backgrounds}
\usetikzlibrary{calc}
\usetikzlibrary{decorations.pathmorphing}
\usetikzlibrary{fit}
\usetikzlibrary{petri}
\usetikzlibrary{plotmarks}
\usetikzlibrary{positioning}
\usetikzlibrary{quotes}
\usetikzlibrary{shapes.misc}
\usetikzlibrary {shapes.multipart}
\usetikzlibrary{spy}
\usepackage{xinttools}
\usetikzlibrary{math}

\usepackage{pgfplots}
\pgfplotsset{compat=newest}
\usepgfplotslibrary{patchplots}
\usepgfplotslibrary{fillbetween}
\usepgfplotslibrary{units}
\usepgfplotslibrary{polar}

\usepgfplotslibrary{external}
\usepackage{siunitx}

\tikzset{
    every node/.append style={font=\footnotesize},
    every label/.append style={font=\footnotesize}
}
\pgfplotsset{
    every axis legend/.append style={
        fill opacity=0.8,
        draw=none,
    },
    every axis/.append style={
        x grid style={darkgray!60},
        xmajorgrids,
        y grid style={darkgray!60},
        ymajorgrids,
    }
}

\usepackage[font=small]{caption}
\usepackage{subcaption}

\usepackage{amsmath,amssymb,amsfonts}
\usepackage{algorithmic}

\usepackage{xifthen}
\newcommand{\prob}[1][]{%
\ifthenelse{\isempty{#1}}%
      {\ensuremath{P}}%
    {\ensuremath{P\left\(#1\right\)}}%
}

\usepackage{amsthm}
\theoremstyle{remark}

    \usepackage{xspace}
\newcommand{\mmwave}{mmWave\xspace}
\newcommand{\subthz}{sub-THz\xspace}
\newcommand{\Subthz}{Sub-THz\xspace}
\newcommand{\subghz}{sub-\SI{10}{\giga\hertz}\xspace}

\newcommand{\update}[1]{#1}
\definecolor{direct}{HTML}{26A269}
\definecolor{pmf}{HTML}{800000}

\newacronym{2d}{2D}{two-dimensional}
\newacronym{3d}{3D}{three-dimensional}
\newacronym{3gpp}{3GPP}{3rd generation partnership project}
\newacronym{4g}{4G}{fourth-generation}
\newacronym{5g}{5G}{fifth-generation}
\newacronym{6dof}{6DoF}{six degrees of freedom}
\newacronym{6g}{6G}{sixth-generation}
\newacronym{abp}{ABP}{authentication by personalisation}
\newacronym{ac}{AC}{alternating current}
\newacronym{aclr}{ACLR}{adjacent channel leakage ratio}
\newacronym{adc}{ADC}{analog-to-digital converter}
\newacronym{adr}{ADR}{adaptive data rate}
\newacronym{aes}{AES}{Advanced Encryption Standard}
\newacronym{afe}{AFE}{analog front end}
\newacronym{ag}{AG}{array gain}
\newacronym{agps}{A-GPS}{Assisted Global Positioning System} %
\newacronym{ai}{AI}{artificial intelligence}
\newacronym{alk}{ALK}{Alkaline}
\newacronym{am}{AM}{amplitude modulation}
\newacronym{amam}{AM/AM}{amplitude modulation to amplitude modulation}
\newacronym{amc}{AMC}{automatic modulation classification}
\newacronym{amp}{AMP}{approximate message passing}
\newacronym{ampm}{AM/PM}{amplitude modulation to phase modulation}
\newacronym{aoa}{AOA}{angle-of-arrival}
\newacronym{aod}{AOD}{angle-of-departure}
\newacronym{ap}{AP}{access point}
\newacronym{api}{API}{application program interface}
\newacronym{apt}{APT}{acoustic power transfer}
\newacronym{apu}{APU}{access point unit}
\newacronym{ar}{AR}{augmented reality}
\newacronym{arima}{ARIMA}{Auto-Regressive Integrated Moving Average}
\newacronym{arp}{ARP}{Antenna Reference Point}
\newacronym{asic}{ASIC}{application specific integrated circuit}
\newacronym{ask}{ASK}{amplitude-shift keying}
\newacronym{at}{AT}{ATtention}
\newacronym{auv}{AUV}{autonomous underwater vehicle}
\newacronym{awg}{AWG}{arbitrary waveform generator}
\newacronym{awgn}{AWGN}{additive white Gaussian noise}
\newacronym{baw}{BAW}{bulk acoustic wave}
\newacronym{bb}{BB}{base-band}
\newacronym{bc}{BC}{backscatter communication}
\newacronym{bd}{BD}{backscatter device}
\newacronym{be}{BE}{Belgium}
\newacronym{ber}{BER}{bit error rate}
\newacronym{bf}{BF}{beamforming}
\newacronym{bga}{BGA}{ball grid array}
\newacronym{bldc}{BLDC}{brushless DC}
\newacronym{bod}{BOD}{Brown-Out Detection}
\newacronym{bom}{BOM}{bill of materials}
\newacronym{bpsk}{BPSK}{binary phase-shift keying}
\newacronym{bs}{BS}{base station}
\newacronym{bw}{BW}{bandwidth}
\newacronym{ca}{CA}{carrier emitter}
\newacronym{cad}{CAD}{channel activity detection}
\newacronym{cars}{CARS}{calibration reference signal}
\newacronym{cbm}{CBM}{condition based maintenance}
\newacronym{cc}{CC}{constant current}
\newacronym{ccdf}{CCDF}{complementary cumulative distribution function}
\newacronym{ccnn}{CCNN}{circular convolutional neural network}
\newacronym{ccs}{CCS}{correlative channel sounder}
\newacronym{cdf}{CDF}{cumulative distribution function}
\newacronym{cdma}{CDMA}{code division-multiple access}
\newacronym{cdrx}{CDRX}{connected mode DRX}
\newacronym{ce}{CE}{coverage enhancement}
\newacronym{ced}{CED}{cumulative energy density}
\newacronym{cf}{CF}{cell-free}
\newacronym{cfmmimo}{CF-mMIMO}{cell-free massive MIMO}
\newacronym{cfo}{CFO}{carrier frequency offset}
\newacronym{cir}{CIR}{channel impulse response}
\newacronym{cla}{CLA}{closed-loop approach}
\newacronym{clk}{CLK}{clock}
\newacronym{cmos}{CMOS}{complementary metal oxide semiconductor}
\newacronym{cost}{COST}{commercial off-the-shelf}
\newacronym{cots}{COTS}{commercial off-the-shelf}
\newacronym{cp}{CP}{cyclic prefix}
\newacronym{cpt}{CPT}{capacitive power transfer}
\newacronym{cpu}{CPU}{central-processing unit}
\newacronym{cpw}{CPW}{coplanar waveguide}
\newacronym{cqi}{CQI}{channel quality indicator}
\newacronym{cr}{CR}{coding rate}
\newacronym{crc}{CRC}{cyclic redundancy check}
\newacronym{crlb}{CRLB}{Cram\'er-Rao lower bound}
\newacronym{crs}{CRS}{cell reference signal}
\newacronym{cs}{CS}{compressed sensing}
\newacronym{cse}{CSE}{channel state estimation}
\newacronym{csi}{CSI}{channel state information}
\newacronym{csp}{CSP}{contact service point}
\newacronym{css}{CSS}{chirp spread spectrum}
\newacronym{cu}{CU}{central unit}
\newacronym{cv}{CV}{constant voltage}
\newacronym{cw}{CW}{continuous wave}
\newacronym{d2d}{D2D}{device-to-device}
\newacronym{dab}{DAB}{Digital Audio Broadcasting}
\newacronym{dac}{DAC}{digital-to-analog converter}
\newacronym{daq}{DAQ}{data acquisition system}
\newacronym{das}{DAS}{distributed antenna systems}
\newacronym{dbpsk}{DBPSK}{Differential Binary Phase Shift Keying}
\newacronym{dc}{DC}{direct current}
\newacronym{dcc}{DCC}{dynamic cooperation clustering}
\newacronym{ddc}{DDC}{digital down conversion}
\newacronym{de}{DE}{drain efficiency}
\newacronym{dft}{DFT}{discrete Fourier transform}
\newacronym{dl}{DL}{downlink}
\newacronym{dlc}{DLC}{Distributed Laser Charging}
\newacronym{dli}{DLI}{direct link interference}
\newacronym{dlt}{DLT}{Distributed Ledger Technology}
\newacronym{dm}{DM}{diffuse multipath}
\newacronym{dma}{DMA}{Direct Memory Access}
\newacronym{dmac}{DMAC}{Direct Memory Access Controller}
\newacronym{dmc}{DMC}{diffuse multipath component}
\newacronym{dmimo}{D-MIMO}{distributed MIMO}
\newacronym{doa}{DOA}{direction-of-arrival}
\newacronym{dp}{DP}{differencial privacy}
\newacronym{dpd}{DPD}{digital pre-distortion}
\newacronym{dpdk}{DPDK}{Data Plane Development Kit}
\newacronym{dram}{DRAM}{dynamic random-access memory}
\newacronym{drcs}{$\Delta$RCS}{differential-radar cross section}
\newacronym{drx}{DRX}{Discontinuous Reception Mode}
\newacronym{dsb}{DSB}{double-sideband}
\newacronym{dsl}{DSL}{digital subscriber line}
\newacronym{dsp}{DSP}{digital signal processing}
\newacronym{dss}{DSS}{Dataset Storage Standard}
\newacronym{duc}{DUC}{digital up-converter}
\newacronym{dvb}{DVB}{Digital Video Broadcasting}
\newacronym{e2e}{E2E}{end-to-end}
\newacronym{easa}{EASA}{European Union Aviation Safety Agency}
\newacronym{ebg}{EBG}{electromagnetic bandgap}
\newacronym{ec}{EC}{European Commission}
\newacronym{ecc}{ECC}{elliptic curve cryptography}
\newacronym{ecdf}{eCDF}{empirical cumulative distribution function}
\newacronym{ecdlp}{ECDLP}{elliptic curve discrete logarithm problem}
\newacronym{ecsp}{ECSP}{edge computing service point}
\newacronym{edlc}{EDLC}{electrostatic double-layer capacitors}
\newacronym{edrx}{eDRX}{Extended Discontinuous Reception Mode}
\newacronym{ee}{EE}{energy efficiency}
\newacronym{egprs}{EGPRS}{Enhanced Data Rates for GSM Evolution}
\newacronym{eh}{EH}{energy harvesting}
\newacronym{eirp}{EIRP}{equivalent isotropically radiated power}
\newacronym{em}{EM}{electromagnetic}
\newacronym{embb}{eMBB}{enhanced Mobile Broadband}
\newacronym{en}{EN}{energy neutral}
\newacronym{end}{END}{energy-neutral device}
\newacronym{enob}{ENOB}{effective number of bits}
\newacronym{eol}{EoL}{end of life}
\newacronym{epu}{EPU}{edge processing unit}
\newacronym{er}{ER}{Energy Receiver}
\newacronym{erp}{ERP}{effective radiation power}
\newacronym[plural=ESCs,firstplural=electronic speed controllers (ESCs)]{esc}{ESC}{electronic speed control}
\newacronym{esd}{ESD}{electrostatic discharge}
\newacronym{esl}{ESL}{electronic shelf label}
\newacronym{esr}{ESR}{equivalent series resistance}
\newacronym{et}{ET}{Energy Transmitter}
\newacronym{etsi}{ETSI}{European Telecommunications Standards Institute}
\newacronym{evd}{EVD}{eigenvalue decomposition}
\newacronym{evm}{EVM}{Error Vector Magnitude}
\newacronym{ewlb}{eWLB}{Embedded Wafer Level Ball Grid Array}
\newacronym{fa}{FA}{federation anchor}
\newacronym{fair}{FAIR}{Findability, Accessibility, Interoperability, and Reuse of digital assets}
\newacronym{fd}{FD}{front-haul distance}
\newacronym{fde}{FDE}{frequency domain equalizer}
\newacronym{fdm}{FDM}{frequency-division multiplexing}
\newacronym{fdma}{FDMA}{frequency division multiple access}
\newacronym{fem}{FEM}{finite element analysis}
\newacronym{fembb}{feMBB}{further enhanced mobile broadband}
\newacronym{fft}{FFT}{fast Fourier transform}
\newacronym{fh}{FH}{fronthaul}
\newacronym{fhss}{FHSS}{frequency hopping spread spectrum}
\newacronym{fifo}{FIFO}{First In, First Out}
\newacronym{fim}{FIM}{Fisher information matrix}
\newacronym{fits}{FITS}{Flexible Image Transport System}
\newacronym{fl}{FL}{federated learning}
\newacronym{fom}{FoM}{figure of merit}
\newacronym{fpga}{FPGA}{field-programmable gate array}
\newacronym{fram}{FRAM}{Ferroelectric Random Access Memory}
\newacronym{fsk}{FSK}{frequency shift keying}
\newacronym{fss}{FSS}{frequency selective surface}
\newacronym{fwa}{FWA}{Fixed Wireless Access}
\newacronym{gb}{GB}{grant-based}
\newacronym{gdpr}{GDPR}{general data protection regulation}
\newacronym{gf}{GF}{grant-free}
\newacronym{gmsk}{GMSK}{Gaussian minimum-shift keying}
\newacronym{gnb}{gNB}{Next Generation Node B}
\newacronym{gni}{GNI}{gross national income}
\newacronym{gnn}{GNN}{graph neural network}
\newacronym{gnss}{GNSS}{global navigation satellite system}
\newacronym{gpclk}{GPCLK}{general purpose clock}
\newacronym{gpio}{GPIO}{General-Purpose Input/Output}
\newacronym{gpl}{GPL}{GNU General Public License}
\newacronym{gprs}{GPRS}{General Packet Radio Services}
\newacronym{gps}{GPS}{Global Positioning System}
\newacronym{gpu}{GPU}{graphical processing unit}
\newacronym{grc}{GRC}{GNU Radio Companion}
\newacronym{gscm}{GSCM}{geometry‐based stochastic model}
\newacronym{gsm}{GSM}{Global System for Mobile Communications}  %
\newacronym{gwp}{GWP}{Global Warming Potential}
\newacronym{harq}{HARQ}{hybrid automatic repeat request}
\newacronym{hat}{HAT}{hardware attached on top}
\newacronym{hcs}{HCS}{human-centric services}
\newacronym{hdf5}{HDF5}{Hierarchical Data Format version 5}
\newacronym{hdi}{HDI}{High-density Interconnect} 
\newacronym{hfss}{HFSS}{High Frequency Simulator Software}
\newacronym{hmd}{HMD}{head-mounted display}
\newacronym{hpbm}{HPBM}{half power beam width}
\newacronym{HyMPRo}{HyMPRo}{Hybrid Multi-Path Routing algorithm}
\newacronym{i.i.d.}{i.i.d.}{independent and identically distributed}
\newacronym{i2c}{I2C}{Inter-Integrated Circuit}
\newacronym{iaq}{IAQ}{Indoor Air Quality}
\newacronym{ib}{IB}{in-band}
\newacronym{ibo}{IBO}{input back-off}
\newacronym{ic}{IC}{integrated circuit}
\newacronym{ici}{ICI}{intercarrier interference}
\newacronym{icnirp}{ICNIRP}{International Commission on Non-Ionizing Radiation Protection}
\newacronym{id}{ID}{information decoding}
\newacronym{idft}{IDFT}{inverse discrete Fourier transform}
\newacronym{if}{IF}{intermediate-frequency}
\newacronym{iid}{i.i.d.}{independently and identically distributed}
\newacronym{iis}{IIS}{integrated information system}
\newacronym{im}{IM}{intermodulation}
\newacronym{imd}{IMD}{intermodulation distortion}
\newacronym{imu}{IMU}{inertial measurement unit}
\newacronym{io}{IO}{input/output}
\newacronym{ioe}{IoE}{Internet of Everything}
\newacronym{iot}{IoT}{Internet of Things}
\newacronym{ipt}{IPT}{inductive power transfer}
\newacronym{ipy}{IPY}{Interventions per Year}
\newacronym{iq}{IQ}{in-phase and quadrature}
\newacronym{iqi}{IQI}{IQ imbalance}
\newacronym{ir}{IR}{infrared}
\newacronym{isi}{ISI}{intersymbol interference}
\newacronym{ism}{ISM}{industrial, scientific and medical}
\newacronym{isp}{ISP}{internet service provider}
\newacronym{jesd}{JESD}{Joint Electron Devices Engineering Council}
\newacronym{jfet}{JFET}{junction field effect transistor}
\newacronym{kpi}{KPI}{key performance indicator}
\newacronym{kvi}{KVI}{key value indicator}
\newacronym{larva}{LARVA}{LARge Virtual Array}
\newacronym{lca}{LCA}{life cycle assessment}
\newacronym{lco}{LCO}{lithium cobalt oxide}
\newacronym{ldo}{LDO}{Low-dropout voltage regulator}
\newacronym{ldpc}{LDPC}{low-density parity-check}
\newacronym{led}{LED}{Light Emitting Diode}
\newacronym{less}{LESS}{Low Energy Scheduler Solution}
\newacronym{lfp}{LFP}{lithium iron phosphate}
\newacronym{lib}{LIB}{Lithium-Ion Battery}
\newacronym{lic}{LIC}{lithium-ion capacitor}
\newacronym{lid}{LID}{Lithium Iron Disulfide}
\newacronym{lidar}{LiDAR}{light detection and ranging}
\newacronym{liion}{Li-ion}{lithium-ion}
\newacronym{lipo}{LiPo}{lithium polymer}
\newacronym{lis}{LIS}{large intelligent surface}
\newacronym{llh}{LLH}{log-likelihood}
\newacronym{lmd}{LMD}{Lithium Manganese Dioxide}
\newacronym{lmmse}{LMMSE}{least minimum mean square error}
\newacronym{lmo}{LMO}{lithium ion manganese oxide}
\newacronym{lna}{LNA}{low-noise amplifier}
\newacronym{lo}{LO}{local oscillator}
\newacronym{lora}{LoRa}{long range}
\newacronym{lorawan}{LoRaWAN}{long-range wide-area network}
\newacronym{los}{LoS}{line-of-sight}
\newacronym{lp}{LP}{linear programming}
\newacronym{lpf}{LPF}{low-pass filter}
\newacronym{lpt}{LPT}{laser power transfer}
\newacronym{lpwa}{LPWA}{Low Power Wide Area}
\newacronym{lpwan}{LPWAN}{low-power wide-area network}
\newacronym{lpwans}{LPWANs}{Low-Power Wide-Area Networks}
\newacronym{lqi}{LQI}{link quality indicator}
\newacronym{lrelu}{LReLU}{leaky rectified linear unit}
\newacronym{lrt}{LRT}{likelihood-ratio test}
\newacronym{ls}{LS}{least squares}
\newacronym{lsa}{LSA}{large synthetic array}
\newacronym{lsf}{LSF}{large-scale fading}
\newacronym{lsfc}{LSFC}{large-scale fading component}
\newacronym{lstm}{LSTM}{Long Short-Term Memory}
\newacronym{ltc}{LTC}{lithium thionyl chloride}
\newacronym{lte}{LTE}{Long Term Evolution}
\newacronym{lti}{LTI}{linear time-invariant}
\newacronym{lto}{LTO}{lithium titanate}
\newacronym{lusta}{LUSTA 5G }{Logistique mUltimodale Sécuritaire Téléopérée \& Autonome 5G}
\newacronym{m2m}{M2M}{machine to machine}
\newacronym{mac}{MAC}{Medium Access Control}
\newacronym{mate}{MATE}{millimeter-wave MIMO testbed}
\newacronym{mc}{MC}{Monte Carlo}
\newacronym{mcl}{MCL}{Maximum Coupling Loss}
\newacronym{mcs}{MCS}{modulation and coding scheme}
\newacronym{mcu}{MCU}{microcontroller unit}
\newacronym{mec}{MEC}{multi-access edge computing}
\newacronym{mems}{MEMS}{micro-electromechanical systems}
\newacronym{mf}{MF}{matched filter}
\newacronym{mimo}{MIMO}{multiple-input multiple-output}
\newacronym{miso}{MISO}{multiple-input single-output}
\newacronym{ml}{ML}{machine learning}
\newacronym{mlp}{MLP}{multilayer perceptron}
\newacronym{mmic}{MMIC}{monolithic microwave integrated circuit}
\newacronym{mmimo}{mMIMO}{massive MIMO}
\newacronym{mmse}{MMSE}{minimum mean square error}
\newacronym{mmtc}{mMTC}{massive machine-typed communication}
\newacronym{mmwave}{mmWave}{millimeter-wave}
\newacronym{mn}{MN}{matching network}
\newacronym{mosfet}{MOSFET}{metal-oxide semiconductor field effect transistor}
\newacronym{mpc}{MPC}{multipath component}
\newacronym{mppt}{MPPT}{maximum power point tracking}
\newacronym{mr}{MR}{maximum ratio}
\newacronym{mrc}{MRC}{maximum ratio combining}
\newacronym{mrc_em}{MRC}{maximum ratio combining}
\newacronym{mrc_EM}{MRC}{Magnetic Resonance Coupling}
\newacronym{mrt}{MRT}{maximum ratio transmission}
\newacronym{mse}{MSE}{mean square error}
\newacronym{msk}{MSK}{Minimum-Shift Keying}
\newacronym{mtc}{MTC}{Machine-Type Communication}
\newacronym{multi-rat}{Multi-RAT}{multiple radio access technology}
\newacronym{multirat}{Multi-RAT}{Multiple Radio Access Technology}
\newacronym{music}{MUSIC}{MUltiple SIgnal Classification}
\newacronym{navauwall}{NAVAUWALL}{AUtomated NAVigation in WALLonia}
\newacronym{nb}{NB}{narrowband}
\newacronym{nbiot}{NB-IoT}{narrowband IoT}
\newacronym{nca}{NCA}{nickel cobalt aluminum}
\newacronym{netcdf}{NetCDF}{Network Common Data Form}
\newacronym{nf}{NF}{noise figure}
\newacronym{nfc}{NFC}{near-field communication}
\newacronym{nfv}{NFV}{network function virtualization}
\newacronym{ngmn}{NGMN}{Next Generation Mobile Networks }
\newacronym{ni}{NI}{National Instruments}
\newacronym{nicd}{NiCd}{nikkel cadmium}
\newacronym{nimh}{NiMH}{nikkel metal hydride}
\newacronym{nlos}{NLoS}{non-line-of-sight}
\newacronym{nmc}{NMC}{nickel manganese cobalt}
\newacronym{nmos}{nMOS}{n-channel metal-oxide semiconductor}
\newacronym{nn}{NN}{neural network}
\newacronym{nnls}{NNLS}{non-negative least squares}
\newacronym{noma}{NOMA}{non-orthogonal multiple access}
\newacronym{np}{NP}{Neyman-Pearson}
\newacronym{npbch}{NPBCH}{Narrowband Physical Broadcast Channel}
\newacronym{npss}{NPSS}{Narrow Band Primay Synchronization Signal}
\newacronym{nr}{NR}{New Radio}
\newacronym{nrs}{NRS}{Narrow Band Reference Signal}
\newacronym{nsss}{NSSS}{Narrowband Secondary Synchronization Signal}
\newacronym{ntp}{NTP}{network time protocol}
\newacronym{oai}{OAI}{OpenAirInterface} %
\newacronym{obw}{OBW}{occupied bandwidth}
\newacronym{ofdm}{OFDM}{orthogonal frequency-division multiplexing}
\newacronym{ofdma}{OFDMA}{orthogonal frequency-division multiple access}
\newacronym{ofdmim}{OFDM-IM}{OFDM with index modulation}
\newacronym{oma}{OMA}{orthogonal multiple access}
\newacronym{oob}{OOB}{out-of-band}
\newacronym{ook}{OOK}{on-off keying}
\newacronym{oran}{O-RAN}{open radio-access network}
\newacronym{os}{OS}{operating system}
\newacronym{ota}{OTA}{over-the-air}
\newacronym{otaa}{OTAA}{over-the-air authentication}
\newacronym{p1}{P1}{Phase 1}
\newacronym{p2}{P2}{Phase 2}
\newacronym{p2p}{P2P}{point-to-point}
\newacronym{pa}{PA}{power amplifier}
\newacronym{pae}{PAE}{power-added efficiency}
\newacronym{pana}{PanA}{Panel A}
\newacronym{panb}{PanB}{Panel B}
\newacronym{papr}{PAPR}{peak-to-average power ratio}
\newacronym{pc}{PC}{pilot count}
\newacronym{pcb}{PCB}{printed circuit board}
\newacronym{pcg}{PCG}{power consumption gain}
\newacronym{pcie}{PCIe}{Peripheral Component Interconnect Express}
\newacronym{pcsi}{PCSI}{perfect channel state information}
\newacronym{pd}{PD}{powered device}
\newacronym{pdcch}{PDCCH}{physical downlink control channel}
\newacronym{pdf}{PDF}{probability density function}
\newacronym{pdp}{PDP}{power delay profile}
\newacronym{pdsch}{PDSCH}{physical downlink shared channel}
\newacronym{pe}{PE}{processing element}
\newacronym{peb}{PEB}{positioning error bound}
\newacronym{per}{PER}{packet error rate}
\newacronym{pet}{PET}{privacy enhancing technology}
\newacronym{pg}{PG}{path gain}
\newacronym{pgd}{PGD}{proximal gradient descent}
\newacronym{phy}{PHY}{physical}
\newacronym{pki}{PKI}{public key infrastucture}
\newacronym{pl}{PL}{path loss}
\newacronym{pll}{PLL}{phase-locked loop}
\newacronym{pmf}{PMF}{polymer microwave fiber}
\newacronym{pmu}{PMU}{Power Management Unit}
\newacronym{pn}{PN}{pseudo-noise}
\newacronym{poe}{PoE}{power-over-Ethernet}
\newacronym{pps}{1PPS}{pulse per second}
\newacronym{pr}{PR}{phase reversal}
\newacronym{prbs}{PRBs}{Physical Resource Blocks}
\newacronym{prs}{PRS}{Peripheral Reflex System}
\newacronym{ps}{PS}{Processing System}
\newacronym{psd}{PSD}{power spectral density}
\newacronym{pse}{PSE}{power sourcing equipment}
\newacronym{psk}{PSK}{phase shift keying}
\newacronym{psm}{PSM}{power saving mode}
\newacronym{pss}{PSS}{primary synchronisation signal}
\newacronym{ptp}{PTP}{precision-time protocol}
\newacronym{ptrs}{PTRS}{Phase-Tracking Reference Signals}
\newacronym{ptw}{PTW}{paging time window}
\newacronym{pv}{PV}{photovoltaic}
\newacronym{pw}{PW}{planar wavefront}
\newacronym{pwm}{PWM}{pulse width modulation}
\newacronym{qam}{QAM}{quadrature amplitude modulation}
\newacronym{qos}{QoS}{quality-of-service}
\newacronym{qpsk}{QPSK}{quadrature phase-shift keying}
\newacronym{qrrls}{QR-RLS}{QR decomposition based recursive least squares}
\newacronym{quadriga}{QuaDRiGa}{QUAsi Deterministic RadIo channel GenerAtor}
\newacronym{ra}{RA}{Random Access}
\newacronym{ram}{RAM}{random-access memory}
\newacronym{ran}{RAN}{radio access network}
\newacronym{rar}{RAR}{Random Access Response}
\newacronym{rat}{RAT}{radio access technology}
\newacronym{rb}{Rb}{Rubidium}
\newacronym{ru}{RU}{radio unit}
\newacronym{rbs}{RBS}{radio base station}
\newacronym{rbw}{RBW}{resolution bandwidth}
\newacronym{rcs}{RCS}{radar cross section}
\newacronym{rdl}{RDL}{redistribution layer}
\newacronym{re}{RE}{radio element}
\newacronym{relu}{ReLU}{rectified linear unit}
\newacronym{rf}{RF}{radio frequency}
\newacronym{rfeh}{RFEH}{radio frequency energy harvesting}
\newacronym{rfic}{RFIC}{radio-frequency integrated circuit}
\newacronym{rfid}{RFID}{radio frequency identification}
\newacronym{rfpt}{RFPT}{radio frequency power transfer}
\newacronym{rfsoc}{RFSoC}{Radio Frequency System-on-Chip}
\newacronym{rir}{RIR}{room impulse response}
\newacronym{ris}{RIS}{reflective intelligent surface}%
\newacronym{rllmtc}{RLLMTC}{reliable low latency machine type communication}
\newacronym{rls}{RLS}{recursive least squares}
\newacronym{rms}{RMS}{root-mean-square}
\newacronym{rmse}{RMSE}{root-mean-square error}
\newacronym{rmt}{RMT}{random matrix theory}
\newacronym{rof}{RoF}{radio-over-fiber}
\newacronym{ros}{ROS}{robot operating system}
\newacronym{rpi}{RPi}{Raspberry Pi}
\newacronym{rrc}{RRC}{Radio Resource Connection}
\newacronym{rreq}{RREQ}{route request packet}
\newacronym{rsrp}{RSRP}{Reference Signals Received Power}
\newacronym{rsrq}{RSRQ}{Reference Signal Received Quality}
\newacronym{rss}{RSS}{received signal strength}
\newacronym{rssi}{RSSI}{received signal strength indicator}
\newacronym{rtc}{RTC}{real time clock}
\newacronym{rtf}{RTF}{reader talks first}
\newacronym{rtk}{RTK}{real time kinematics}
\newacronym{rts}{RTS}{ray tracing simulator}
\newacronym{rv}{RV}{random variable}
\newacronym{rw}{RW}{RadioWeaves}
\newacronym{rx}{RX}{receiver}
\newacronym{rzf}{RZF}{regularized zero forcing}
\newacronym{s-parameter}{S-parameter}{scattering parameter}
\newacronym{sa}{SA}{synchronization anchor}
\newacronym{sa5}{SA}{Stand Alone}
\newacronym{sar}{SAR}{specific absorption rate}
\newacronym{sbl}{SBL}{sparse Bayesian learning}
\newacronym{scfdma}{SCFDMA}{single-carrier frequency division multiple access}
\newacronym{sdg}{SDG}{Sustainable Development Goal}
\newacronym{sdm}{SDM}{sigma-delta modulator}
\newacronym{sdma}{SDMA}{spatial-division multiple access}
\newacronym{sdn}{SDN}{software-defined network}
\newacronym{sdof}{SDoF}{sigma-delta over fiber}
\newacronym{sdr}{SDR}{software-defined radio}
\newacronym{se}{SE}{spectral efficiency}
\newacronym{sei}{SEI}{specific emitter identification}
\newacronym{ser}{SER}{symbol-error rate}
\newacronym{sf}{SF}{spreading factor}
\newacronym{sfn}{SFN}{single frequency network}
\newacronym{sfp}{SFP}{small form-factor pluggable}
\newacronym{sha}{SHA}{Secure Hash Algorithm}
\newacronym{SigMF}{SigMF}{Signal Metadata Format}
\newacronym{simo}{SIMO}{single-input multiple-output}
\newacronym{sinr}{SINR}{signal-to-interference-plus-noise ratio}
\newacronym{siso}{SISO}{single-input single-output}
\newacronym{slam}{SLAM}{simultaneous localization and mapping}
\newacronym{slc}{SLC}{spatial leakage suppression}
\newacronym{slerp}{SLERP}{spherical linear interpolation}
\newacronym{sma}{SMA}{SubMiniature version A}
\newacronym{smc}{SMC}{specular multipath component}
\newacronym{smps}{SMPS}{switched mode power supply}
\newacronym{sndr}{SNDR}{signal-to-noise-and-distortion ratio}
\newacronym{snidr}{SNIDR}{signal-to-noise-and-interference-and-distortion ratio}
\newacronym{snir}{SNIR}{signal-to-interference-plus-noise ratio}
\newacronym{snr}{SNR}{signal-to-noise ratio}
\newacronym{soc}{SoC}{state of charge}
\newacronym{SoC}{SOC}{System on Chip}
\newacronym{sp}{SP}{service point}
\newacronym{spdt}{SPDT}{single pole double throw}
\newacronym{spi}{SPI}{Serial Peripheral Interface}
\newacronym{spst}{SPST}{single pole single throw}
\newacronym{sram}{SRAM}{static random-access memory}
\newacronym{srd}{SRD}{short-range device}
\newacronym{srls}{SRLS}{standard recursive least squares}
\newacronym{srs}{SRS}{Sounding Reference Signal}
\newacronym{ssb}{SSB}{synchronisation signal block}
\newacronym{ssd}{SSD}{solid state drive}
\newacronym{ssq}{SSQ}{simulator sickness questionnaire}
\newacronym{steam}{STEAM}{science, technology, engineering, the arts, and mathematics}
\newacronym{svd}{SVD}{singular value decomposition}
\newacronym{sw}{SW}{spherical wavefront}
\newacronym{swipt}{SWIPT}{simultaneous wireless information and power transfer}
\newacronym{synce}{SyncE}{Synchronous Ethernet}
\newacronym{tau}{TAU}{tracking area update}
\newacronym{tcer}{TCER}{transported to consumed energy ratio}
\newacronym{tcp}{TCP}{Transmission Control Protocol}
\newacronym{tcxo}{TCXO}{temperature compensated crystal oscillator}
\newacronym{tdd}{TDD}{time division duplexing}
\newacronym{tdma}{TDMA}{time division-multiple access}
\newacronym{tdoa}{TDOA}{time-difference-of-arrival}
\newacronym{thz}{THz}{Terahertz}
\newacronym{to}{TO}{timing offset}
\newacronym{toa}{TOA}{time-of-arrival}
\newacronym{tof}{ToF}{time-of-flight}
\newacronym{tosm}{TOSM}{through-open-short-match}
\newacronym{tpms}{TPMS}{Tire-Pressure Monitoring System}
\newacronym{trl}{TRL}{technology readyness level}
\newacronym{trp}{TRP}{Transmission Reception Point}
\newacronym{tsn}{TSN}{time-sensitive networking}
\newacronym{ttf}{TTF}{tag talks first}
\newacronym{ttff}{TTFF}{Time To First Fix}
\newacronym{ttm}{TTM}{time to market}
\newacronym{ttn}{TTN}{The Things Network}
\newacronym{tx}{TX}{transmitter}
\newacronym{uart}{UART}{Universal Asynchronous Receiver/Transmitter}
\newacronym{uav}{UAV}{unmanned aerial vehicle}
\newacronym{uc}{UC}{use case}
\newacronym{ucie}{UCIe}{Universal Chiplet Interconnect Express}
\newacronym{udp}{UDP}{User Datagram Protocol}
\newacronym{ue}{UE}{user equipment}
\newacronym{ugv}{UGV}{unmanned ground vehicle}
\newacronym{uhd}{UHD}{USRP hardware driver}
\newacronym{uhf}{UHF}{ultra-high frequency}
\newacronym{ul}{UL}{uplink}
\newacronym{ula}{ULA}{uniform linear array}
\newacronym{uMIMO}{$\mu$-MIMO}{ultra-massive Multiple-Input Multiple-Output}
\newacronym{ummtc}{umMTC}{ultra massive machine type communication}
\newacronym{un}{UN}{United Nations}
\newacronym{upa}{UPA}{uniform planar array}
\newacronym{ura}{URA}{uniform rectangular array}
\newacronym{urllc}{URLLC}{ultra-reliable low-latency communications}
\newacronym{usrp}{USRP}{universal software radio peripheral}
\newacronym{uv}{UV}{unmanned vehicle}
\newacronym{uwb}{UWB}{ultrawideband}
\newacronym{v2v}{V2V}{vehicle-to-vehicle}
\newacronym{vco}{VCO}{voltage-controlled oscillator}
\newacronym{vep}{VEP}{virtual edge platform}
\newacronym{vlc}{VLC}{visible light communication}
\newacronym{vlp}{VLP}{visible light positioning}
\newacronym{vna}{VNA}{vector network analyzer}
\newacronym{voc}{VOC}{Voltatile Organic Compound}
\newacronym{votable}{VOTable}{Virtual Observatory Table}
\newacronym{vr}{VR}{virtual reality}
\newacronym{wb}{WB}{wideband}
\newacronym{wban}{WBAN}{wireless body area network}
\newacronym{wd}{WD}{Wireless distance}
\newacronym{wimax}{WiMAX}{Worldwide Interoperability for Microwave Access}
\newacronym{wlan}{WLAN}{wireless LAN}
\newacronym{wpt}{WPT}{wireless power transfer}
\newacronym{wr}{WR}{White Rabbit}
\newacronym{wrsn}{WRSN}{wireless rechargeable sensor network}
\newacronym{wsn}{WSN}{wireless sensor network}
\newacronym{xets}{XETS}{cross exponentially tapered slot}
\newacronym{xr}{XR}{extended reality}
\newacronym{z3ro}{Z3RO}{zero third-order distortion}
\newacronym{zf}{ZF}{zero-forcing}
\newacronym{zmcscg}{ZMCSCG}{zero mean circularly symmetric complex Gaussian}
\newacronym{ep}{EP}{energy profiler}

\newacronym{hpbw}{HPBW}{half-power beamwidth}

\begin{document}

\title{Distributed Deployment and Dual-Frequency Concepts to Strengthen Sub-THz Wireless Systems}

\author{Liesbet Van der Perre, Gilles Callebaut, Thomas Eriksson, Muris Sarajlic, Christian Fager, Fredrik Tufvesson, Buon Kiong Lau, and Erik G. Larsson%
\thanks{
\textit{Gilles Callebaut and Liesbet Van der Perre are with Department of Electrical Engineering, KU Leuven.}%
 \textit{Erik G. Larsson is with the Department of Electrical Engineering (ISY), Link\"oping University.}
 \textit{Fredrik Tufvesson and Buon Kiong Lau is with the Department of Electrical and Information Technology, Lund University.}
  \textit{Thomas Eriksson is with the Department of Electrical Engineering, Chalmers University of Technology.}
  \textit{Christian Fager is with the Department of Microtechnology and Nanoscience, Chalmers University of Technology.}
 \textit{Muris Sarajlic is with Ericsson Research.}
\\This research was partially funded by 6GTandem, supported by the Smart Networks and Services Joint Undertaking (SNS JU) under the European Union's Horizon Europe research and innovation programme under Grant Agreement No~101096302.
}%
\thanks{For the purpose of open access, the author has applied a CC BY public copyright license to any Author Accepted Manuscript version arising from this submission.}%
}

\maketitle

\begin{abstract}%
    The vast bandwidth available at \subthz frequencies holds great
promise for high-speed wireless access, precise localization, and
advanced sensing applications. However, fundamental physical
constraints and technological limitations make the deployment of
reliable \subthz networks challenging. We propose a new paradigm
for \subthz coverage by transmitting the RF signals over \glspl{pmf}
that interconnect low-complexity \glspl{ru} in a daisy-chain
configuration.  The distributed architecture ensures that \glspl{ue}
connect to \glspl{ru} in their proximity, reducing path loss and
mitigating blocking. The \glspl{ru} leverage low-complexity, compact
integrated antenna modules.  Additionally, dual-frequency \emph{tandem
operation} is proposed, integrating the \subthz system with a \subghz
system that provides control signalling and a robust fallback solution
for the \subthz system. This proposed tandem architecture can open up
the full potential of \subthz technology and paves the way to cost-
and energy-efficient, high-performance, real-time connectivity in
dynamic environments.

\end{abstract}


    \section{Introduction}\glsresetall

\begin{figure}[t!]
    \centering
    \includegraphics[width=0.9\linewidth,trim={6cm 2cm 6cm 0},clip]{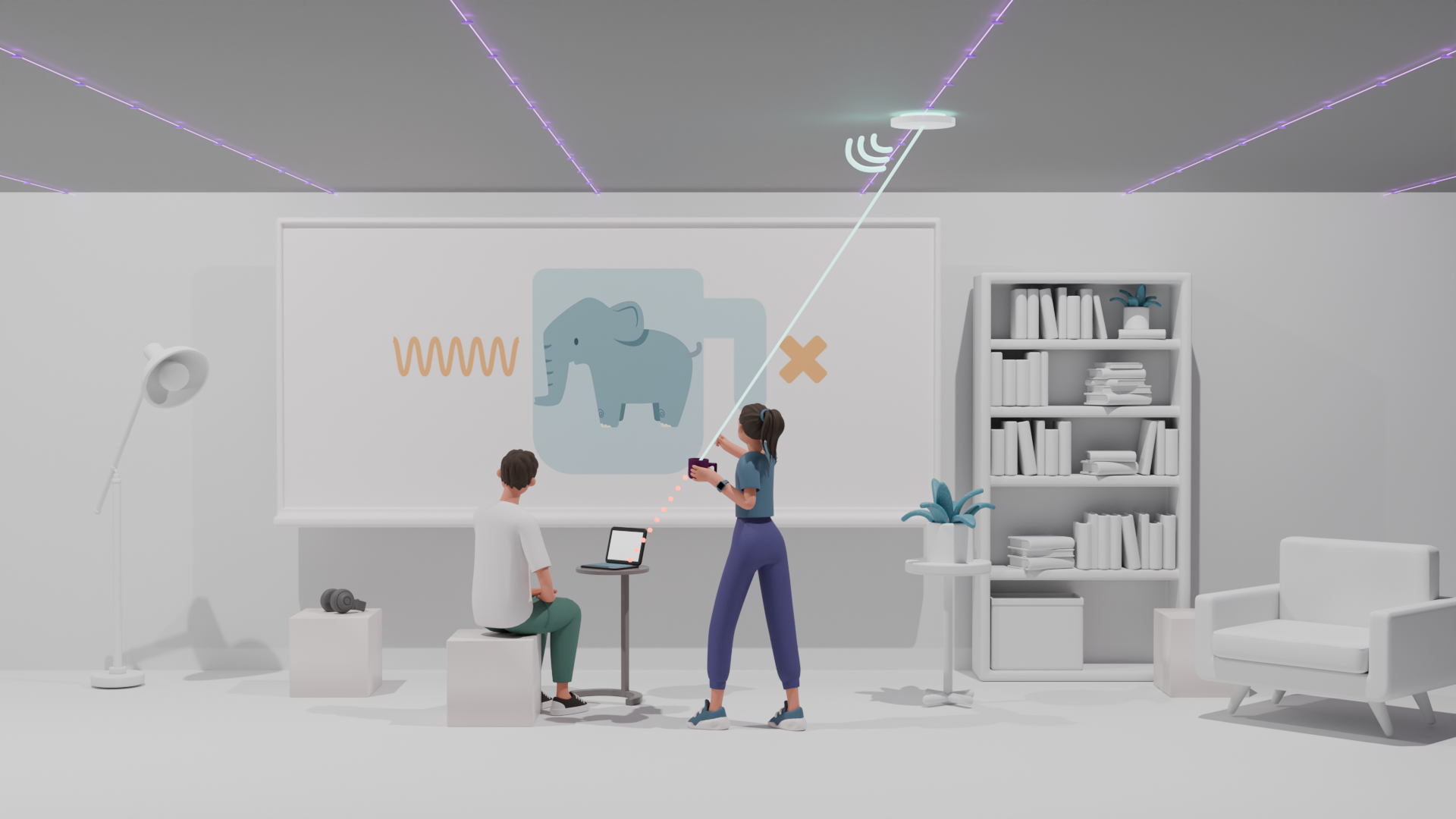}
    \caption{The fragility of \subthz links calls for new deployment approaches. \update{A distributed deployment can strengthen the systems.}}\label{fig:elephant}
\end{figure}

\begin{figure*}[t!]
    \centering
    \includegraphics[width=0.9\linewidth]{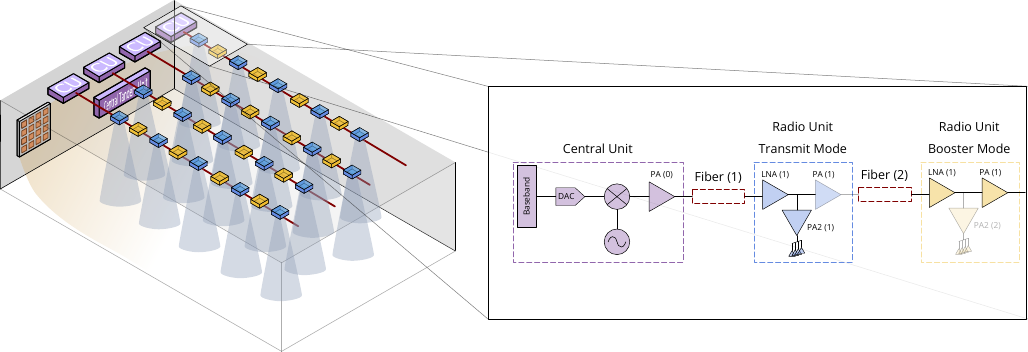}
    \caption{A dense  deployment of distributed low-complexity \subthz
      \glspl{ru} and \subghz with dual-frequency operation. Block diagram of the
      \subthz fiber-based infrastructure (the stripe), consisting of a
      \gls{cu}, and multiple \glspl{ru} connected with
      \glspl{pmf}. \update{As detailed in the close-up, a \gls{ru} can be configured to transmit over the air (blue) or amplify the signal (yellow) to be carried further over the \gls{pmf} to account for the losses.}}\label{fig:Tandem_concepts}
\end{figure*}

Applications such as AR/XR and robotized factories require
\unit{Gbit/s} communication and accurate positioning.
The large bandwidth available at \subthz frequencies suggests
that \update{this band could provide a} ``highway to connectivity heaven''.
This has motivated R\&D teams to progress technology, resulting in \subthz channel
models~\cite{9794668}, hardware modules integrating many antennas and transceivers~\cite{10185423},
and demonstrations of transmission at
\SI{>100}{Gbit/s}~\cite{10443477, 10742620}.  These developments have
demonstrated that very high throughputs can be achieved in \subthz
bands, in particular for static links. However,
providing \update{reliable coverage and} consistent connectivity to non-static terminals at these
frequencies is extremely difficult when compared to operation in \subghz bands.
\Subthz links are fragile. %
The reasons lie in physics of
propagation, \update{blocking and Doppler especially, as explained}
in~\cref{sec:obstacles}, and in \update{losses and intrinsic distortions introduced by}
high-frequency hardware \update{that need to be mitigated}, as elaborated
in~\cref{sec:integration}. \update{Hence,} critical questions remain open: Can
the potential of \subthz be unleashed for applications in need of
reliable links to \update{non-static} users? %
Or will \subthz
systems follow the same path as \mmwave solutions, which in actual
commercial deployments so far remain underused?

\update{Distributed approaches have been proposed to cope with blockage, e.g.~\cite{9353405}, however not considering how the distribution of \subthz signals could be effectuated with a good cost versus signal loss compromise}. In this paper, we present a new perspective on \update{how to provide \subthz connectivity based on low-complexity distribution of \subthz signals with RF-over-\acrlong{pmf} interconnecting compact \acrlongpl{ru} in a daisy-chain configuration (\cref{fig:Tandem_concepts}).} \update{Co-designed dual-frequency transmission and deployment approaches are proposed to mitigate propagation and implementation bottlenecks. The main contributions are}:
\begin{itemize}

      \item A dense deployment of \glspl{ru} \update{in a daisy-chain configuration connected with \glspl{pmf} segments} \update{to bring the \subthz signals close to the terminals, circumventing blocking}. A \gls{ru},
      see \cref{fig:Tandem_concepts}, consists of
      only analog components. This solution
      makes a clean break with existing designs, for example the
      multipoint approach in~\cite{9353405} (cf. \cref{sec:dense}).

    \item Novel radio-over-plastic fiber~\cite{Strömbeck2023icd}\update{, specifically \gls{pmf},}
      technology as a low-complexity and cost-efficient
      solution to distribute the \subthz signals.

    \item Physically compact antennas \update{ distributed over an area} to realize a sufficient link budget.  \update{The arrays integrated in \glspl{ru}, each with only a small number of antennas, together form a sparse large array. Only one or a few \glspl{ru} actively transmit at a given time, creating distinct beams that will "light up" following the allocation of \glspl{ru} to \glspl{ue}.} \glspl{ru} \update{realize rather}
      broad beams that require less \update{frequent updating}
      and enable low-complexity \update{integration}.

    \item Dual-frequency ``tandem'' operation: while
      \subthz band offers very high data throughput and sensing
      resolution, a complementary cooperating lower \subghz band
      offers support on the control plane and as a reliable ``safety
      net''.
\end{itemize}

\section{Propagation and HW implementation: obstacles \update{and mitigating them to reach the  \subthz highway}}\label{sec:obstacles}

Wireless networks have been moving up in frequency
in the quest for more capacity. This has been achieved to a
large extent following the same \update{deployment approach}, i.e., using centrally located
access points with similar \gls{rf} transceiver architectures and
multi-carrier waveforms. While exploiting the same
concepts has been successful for frequencies below \subghz, these approaches are problematic when moving up to the
\mmwave and \subthz bands. As explained below, both radio propagation and hardware
integration at these frequencies are substantially different and more challenging due to
inherent physical limitations. %
\update{The proposed deployment and signalling can mitigate these obstacles, strengthening the links.}

\subsection{Radio propagation at \subthz: a cup of coffee can be the elephant in the room}\label{sec:propagation}

Large-scale and small-scale fading at \subthz impact
performance in a drastically different way as compared to \subghz bands. \update{While the physics are well understood, many reported studies underexpose the resulting impact on communication systems. This is  the proverbial elephant in the room, as  sketched in Fig.~\ref{fig:elephant}.} %
\update{In the following we discuss how the novel deployment and operation concepts address severe propagation effects}.

\textbf{Link budget.}  The effective aperture of an antenna is
inversely proportional to the squared operating frequency. Hence, higher
frequencies result in a smaller effective aperture, leading to
increased loss. Specifically, when comparing transmissions at
\SI{140}{\giga\hertz} and \SI{2.4}{\giga\hertz}, an additional loss of
approximately \SI{30}{\dB} is incurred.  Compensating for
this loss, theoretically, is possible by using an antenna
array with the same physical area, and
thus more directive characteristics \cite{10258328}.  This would in the case above require the integration of \(58 \times 58=3364\) antenna
elements. In practice, the
losses in interconnecting these antennas will counteract the
gain to some extent, as explained
in~\cref{sec:integration}. This many antennas in the array would
result in a \gls{hpbw} of under \SI{2}{degrees} in the broadside, estimated from the array factor. %
\update{Such narrow beams need very frequent tracking for non-static terminals, which is relaxed by the proposed deployment with relatively broad beams (crf.~\cref{sec:mitigation}).} %

\textbf{Shadowing/blocking.} To visualize the different blocking when transmitting at \SI{140}{\giga\hertz} versus
\SI{2.4}{\giga\hertz}, consider the analogy of a coffee mug with a
diameter of \SI{7}{\cm} versus an African elephant that gets \SI{4}{\m} long. These have similar sizes when
expressed in number of wavelengths for the \SI{140}{\giga\hertz} and \SI{2.4}{\giga\hertz} bands, respectively.  Moreover, the liquid
in the cup behaves at \subthz frequencies as an almost perfect
reflector. To illustrate the consequences: suppose someone is holding a cup of coffee at 1~m on the
line to the \gls{ap} from a terminal, equipped with a
\gls{ura} of \num{1024} antennas, as illustrated
in~\cref{fig:elephant}. With a \gls{hpbw} of \SI{7}{degrees}, the cup
effectively becomes like an elephant in the room. This is a ''bad luck'' case, yet for example humans are more
probable to create huge blockers, \update{incurring a  \SI{\sim40}{dB}} loss at \subthz frequencies. %
\update{The proposed distributed deployment can circumvent such blockage, which is a main obstacle to consistent \subthz connectivity.}

\textbf{Sensitivity to mobility.} Doppler is much more severe due
to the extremely short coherence time at \subthz frequencies. Consider
a person walking to the coffee machine at \SI{5}{km/h}. When communicating using \subthz frequencies,
this person will experience a similar Doppler as \subghz communication
does in high-speed trains moving at \SI{300}{km/h}. The impact of high Doppler
is that \update{antenna beams} need to be updated very frequently, in particular
\update{narrow beams created by large antenna arrays}. \update{The proposed deployment, which uses relatively broad beams oriented downward rather than sideways, offers improved resilience to the high Doppler effects characteristic of \subthz frequencies.}

\textbf{Unreliable multipath.} Multipath components create a safety net for communication
links at \subghz frequencies to maintain connectivity when the
\gls{los} is blocked. In \subthz channels, multiple paths are scarce.
Studies~\cite{10438938} reported `it depends' whether this phenomenon
will be significant. This can result in the worst of both worlds:
\glspl{mpc} cannot be relied on for overcoming blocking of the
\gls{los}, nor can one assume that there will be no \glspl{mpc} to reduce equalizer complexity.  A reflection from a wall can
generate a strong \gls{mpc} and be detrimental to the link quality
unless an appropriate equalizer is used. \update{Recent measurements confirm that significant \glspl{mpc} primarily are received via strong side-beams of large arrays.}%

\textit{In conclusion, adopting the conventional \update{central} network
  deployment and transmission \update{schemes} that work well at \subghz is bound to
  lead to unreliable connectivity at \subthz
  frequencies. \update{In contrast, the proposed approach for \subthz systems mitigate the obstacles to provide strong coverage.}}

\subsection{Integration challenges: small components, big losses}\label{sec:integration}

Progress on hardware operating at \subthz frequencies
has been impressive. %
Still, challenges remain significant. \update{Those that are relaxed with the proposed approach are discussed} next.

\textbf{Amplifiers.}  It is difficult to make  \subthz amplifiers
with high output power. At \SI{140}{\giga\hertz},
state-of-the-art amplifiers in commercially viable silicon
technologies can provide $\sim$\SI{20}{dBm} of peak power
~\cite{Wang2020}. This needs to be backed-off by
\SI{5}{dB} \update{to \SI{10}{dB} to transmit 1024-carrier OFDM waveforms. } %
Furthermore,
the efficiency is very low, less than \SI{10}{\percent} under
realistic conditions~\cite{Wang2020}.

\textbf{Phase noise. } The amount of phase noise increases with the
carrier frequency by \SI{20}{dB} per decade. For high-bandwidth
transmission at \subthz, white phase noise will
dominate the distortion. %
Traditional phase tracking techniques cannot mitigate this \update{phase noise, as it is uncorrelated over time.}

\textbf{Antennas and Interconnects. } Low-cost technologies like
\gls{hdi} \gls{pcb} and \gls{ewlb}, which enable practical mass
production, do not (yet) offer low-loss feeding networks
and antenna implementations. For example, in its basic form, the
\gls{rdl} layer at the bottom of an \gls{ewlb} package is used to
implement feeding lines like coplanar waveguides, with insertion
losses on the order of \SI{0.5}{dB/mm}. With a wavelength of
$\sim$\SI{2}{mm}, and the largest dimension of a typical array
element being half a wavelength, the distribution of \gls{rf} signals
to many elements in a large array requires
many mm's of feeding line (and hence many dB's of insertion
loss). Therefore, the array gain from the use of more elements (ideally \SI{3}{dB} for every doubling of
the number of elements) is
significantly degraded by practical feeding of these elements,
potentially even nullifying the extra array gain.

\subsection{\update{Mitigating the obstacles to the \subthz highway}}\label{sec:mitigation}
\update{Adequate system-level designs can mitigate obstacles at \subthz frequencies that make the links fragile. We highlight two key choices that blend well with the proposed deployment approach, further elaborated in~\cref{sec:dense}.}

\textbf{Selecting hardware-friendly waveforms.} \update{Multi-carrier waveforms modulated with higher-order constellations are adopted to offer a high spectral efficiency and cope with multi-path fading, at the expense of a high \gls{papr}. In contrast, we advocate using waveforms requiring low dynamic range that are robust to non-linear distortions
and low-order constellations for \subthz connectivity. These are hardware-friendly and allow relaxation of the requirements of (i) the amplifiers, which can be operated with less back-off; (ii) the oscillators, which inevitably introduce white phase noise; and (iii) the  data converters, whose resolution can be lowered, reducing complexity and energy consumption. The high available bandwidth, realistically considered \SI{20}{GHz} in this paper, can ensure \SI{>10}{Gbit/s} on each \gls{pmf} even with these low-spectral efficienct waveforms. Local network capacity can be scaled up linearly with number of \gls{pmf} stripes to $\sim$~\unit{Tbit/s} in large venues.}

\textbf{Providing coverage with many broad-beams.} %
\update{Many arrays, each with a small number of antennas, can provide stable coverage for non-static users. This approach can cope with both propagation and hardware integration challenges at \subthz: (i) the link budget is less sensitive to user movement thanks to the broader beams and less prone to multi-path as the arrays serve as a spatial filter without high sidelobes, and (ii) the implementation does not suffer from the large interconnect losses coming with arrays with many antenna elements.}

\section{Strengthening \subthz links concept I -- Dense deployments to make friends with physics}\label{sec:dense}

\begin{figure}
\centering
\resizebox{0.95\linewidth}{!}{%
\begin{subfigure}[t]{\linewidth}
    \includegraphics[width=0.985\linewidth]{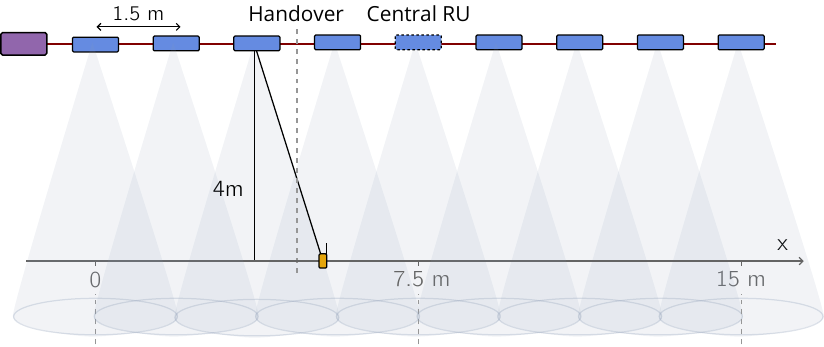}
\end{subfigure}}

\resizebox{0.95\linewidth}{!}{%
\begin{subfigure}[t]{\linewidth}
\centering
\includegraphics[width=0.985\linewidth]{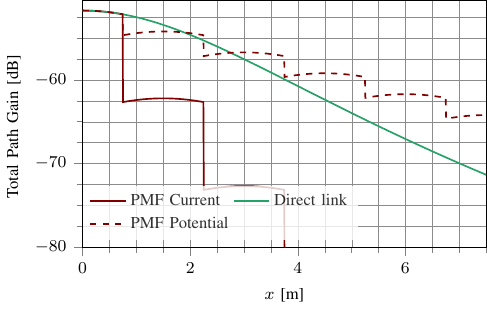}
\end{subfigure}%
}
        \caption{Signal degradation over the air and over the fiber in
          the considered \subthz system. Distributed \glspl{ru}
          outperform a single central \gls{ru} approach, mitigating
          \gls{los} blocking and improving the path gain over the
          air. Every \SI{1.5}{\meter} a \gls{ru} acts as a boost unit
          to amplify the signal. On the fiber, the first couple of
          meters, the signal is noise-limited and transitions to a
          distortion-limited regime after approximately \SI{9}{\meter}.} %
        \label{fig:pl-example}
\end{figure}

\update{The novel paradigm brings the `wireless \subthz entry point' to the proximity of the \glspl{ue} through transmission of the RF signals over \glspl{pmf}, hosting \glspl{ru} connected in a daisy-chain topology. This results in} a distributed, dense deployment of \subthz \glspl{ru} equipped with
antenna arrays with relatively few elements, creating favorable
conditions for providing reliable high-throughput coverage. This architecture, sketched in~\cref{fig:Tandem_concepts}, is reminiscent of
 a spot-based lighting system, where users are
served by the nearest source(s).  Such deployment can overcome propagation obstacles outlined in~\cref{sec:obstacles} by
reducing the link distance, decreasing the risk of blocking,
and lowering the
beam steering complexity and update rate, as also pointed out in \cite{9910182}. The deployment can
ensure that a \gls{ue} at every location is serviced by at least two
\glspl{ru}, as illustrated in~\cref{fig:pl-example}. This enhances
resilience to blocking of the \gls{los} to the nearest \gls{ru}.  The \subthz signals are distributed via \glspl{ru}
over a \acrfull{pmf}. This in contrast to \gls{rof}, where RF signals are converted to optical signals, which are carried over an (optical)
fiber. The \gls{pmf} enables low-complexity
deployment \cite{Strömbeck2023icd} and avoids the need for many
high-frequency transceivers \update{or optical-electrical converters}. However, it also brings limitations as
 no digital \update{baseband} signals are available at the \glspl{ru}.%

For resource allocation, user scheduling, \update{handovers between \glspl{ru}, and minimal on-switching of \subthz hardware,} we propose a dedicated
control channel operating at \subghz that offers reliable
coverage, as explained in~\cref{sec:dual}.  Dual-frequency operation
of networks whereby \mmwave transmission is assisted by lower
frequencies has been investigated before, for example in
\cite{HE2023100610}. However, most developments, such as
Wi-Fi systems, are based on co-located dual-frequency centrally deployed access
points with both carriers optimized for data
transmission. Hence, the \mmwave channel offers a
high-speed data pipe when conditions allow, which in reality turns out
to be rather rarely. In contrast, we propose distinct deployments for
the lower and higher frequencies and aim for \update{consistent} high-data rate
support enabled by the very high bandwidth available at \subthz
frequencies. \update{The inevitable cost overhead of distributed deployment can be justified by the use of \gls{pmf}-based infrastructure, which offers a cost-efficient solution as detailed at the end of this section.}

In the following, we zoom in on the transmission of \subthz
signals over-the-fiber and over-the-air, respectively.

\subsection{Transmitting \subthz signals over PMF}\label{sec:pmf}

\Cref{fig:Tandem_concepts} illustrates the \subthz fiber-based
structure, i.e, the \emph{stripe}. It consists of a central unit,
delivering a \subthz signal on a \gls{pmf}, to a set of \glspl{ru}
that \update{can} transmit the signal over the air. \update{\Glspl{ru} may also be configured to operate in booster mode,} %
with the sole purpose of amplifying the signal along the
fibre, as indicated in \cref{fig:Tandem_concepts}. \update{Disabling a \gls{ru} will disable all subsequent  \glspl{ru}, saving power on the non-activated hardware.}
We \update{restrict the discussion to} the downlink operation here; the uplink is similar but in reverse order.

\textbf{Central unit:} The central unit creates the baseband signals,
and upconverts to \subthz frequencies. Then the signal is amplified in
a power amplifier, and fed into the \gls{pmf}.

\textbf{PMF:} The PMF transports the \subthz signal along the
stripe.  The signal gets attenuated along the PMF, typically in the order of 3~dB per meter and with an additional 3~dB loss
in the chip-to-PMF
transition, i.e., coupler~\cite{Strömbeck2023icd}\cite{10565206}. Possibly, signals
could get spectrally distorted along the way depending on the waveguide design, which may be tailored to avoid this.

\textbf{Radio units:} At chosen intervals, radio units
amplify the signal from the \gls{pmf} and feed to an antenna (or a
phased array). When a \gls{ru} is \update{activated to transmit to a \gls{ue}}, the signal is switched to
the antennas. When the
\gls{ru} is not transmitting or receiving, \update{it acts in booster mode} as
depicted in \cref{fig:Tandem_concepts}.

The signal generation and transmission over the stripe is affected by impairments that can drastically degrade the signal
quality. Two important impairments are \emph{phase noise}, created in
the central unit, and \emph{amplifier nonlinearity} in the \glspl{ru}. Both degrade
the signal quality, and the nonlinearities also lead to
spectral regrowth, potentially breaking spectral mask requirements. A
key limiting factor is the need to restrict the gain of a
\gls{ru} to below $\sim$\SI{30}{dB} to prevent
self-oscillations. These oscillations can occur when signals emitted
from the \gls{ru} are picked up by the \gls{pmf}-to-chip transition
coupler at the input. In practice, maintaining an output-to-input
coupling below 30~dB at these frequencies is extremely challenging.
There is a delicate balance between noise and distortion contributions
along the \gls{pmf}. Feeding a too-strong \gls{rf} signal into the
\gls{pmf} will quickly build up nonlinear intermodulation
distortion. On the other hand, a too-low \gls{rf} signal will  drown in the noise added by each
\gls{ru}. \Cref{fig:pl-example} illustrates a case where the
\gls{rf} signal power has been adjusted to balance signal-to-noise
(S/N) and signal-to-distortion (S/IMD) performance. The signal
distortion in the first part of the \gls{pmf} is dominated by noise. Further along the stripe, the nonlinear distortion becomes the
main limiting factor. Overall, the example demonstrates more than
\SI{30}{dB} signal-to-noise and distortion ratio within the considered
\SI{15}{m} \gls{pmf}.

\textbf{Energy consumption estimation.} \update{
Excellent energy efficiency, $<$\,\SI{10}{pJ/bit}~\cite{10742620} has been achieved for \unit{Gbit/s} wireless \subthz links. However, this is reported for short distances (cm-level) only or with very directive antennas on fixed links, making the comparison with \subghz transmission not entirely fair. %
It is important to acknowledge that the energy consumption of \subthz transmission is impacted by high losses of components and amplifiers, with a typical efficiency $\SI{<10}{\percent}$ in state-of-the-art designs~\cite{Wang2020}. %
In contrast, current \subghz \glspl{pa} reach \SI{>50}{\percent} efficiency.}

\begin{figure}[h]
    \centering

    \resizebox{0.95\linewidth}{!}{%
\begin{subfigure}[t]{\linewidth}
    \includegraphics[width=0.985\linewidth]{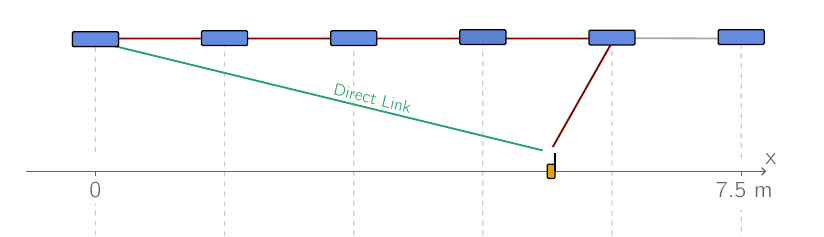}
\end{subfigure}}%

\resizebox{0.95\linewidth}{!}{%
\begin{subfigure}[t]{\linewidth}
\centering
\includegraphics[width=0.985\linewidth]{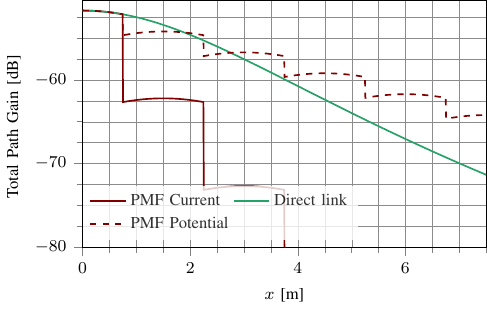}




\end{subfigure}}
\caption{\update{Path gain comparison between a direct over-the-air link (\ref{direct}) and the proposed distributed deployment using \glspl{pmf}.  In the distributed case, the total loss accumulates from both the fiber segments (and couplers) and the over-the-air propagation. Current technology \ref{pmf2} offers a fiber attenuation of \SI{3}{dB/m} and a \SI{3}{dB} loss per coupler, which  experiences significantly path loss. With lower losses  (e.g., a \glspl{pmf} with \SI{1}{dB/m} loss and a \SI{0.5}{dB} coupling loss as in \ref{pmf3}),  the distributed deployment could become more energy-efficient than a central deployment.
        }}
    \label{fig:PG_TOTAL}
\end{figure}

\update{Sub-THz systems must  overcome blocking and shadowing that  induce losses of \SI{40}{dB} and higher. While this problem cannot be solved by increasing the transmit power, our proposed distributed deployment addresses it. However, the solution comes at the expense of energy consumption in \glspl{ru} that need to boost the signals to compensate for losses on the \gls{pmf}. In fully  \gls{los} conditions, a central deployment will consume less energy, as illustrated in \cref{fig:PG_TOTAL} %
, given the current losses in \gls{pmf} fibers.}

\update{Based on efficiency numbers for state-of-the-art amplifiers~\cite{Wang2020}, and assuming \SI{10}{mW} peak output power and class~A operation at \subthz, each \gls{ru}, whether in transmit or in booster mode, is expected to consume
$\sim$\SI{500}{mW}. The total energy consumption will be determined by the \gls{cu}-to-\gls{ue} distance, since inactive \glspl{ru} can be disabled.  Energy consumption scales linearly with stripe length well beyond \SI{100}{m}. Since inactive \glspl{ru} will be disabled, the average distance from the \gls{cu} to the \glspl{ue} determines the energy consumption. The full transceiver functionality in the \gls{cu} could consume as little as \SI{100}{mW} in transmit or receive mode~\cite{10742620}.
This yields a total power consumption of \SI{\sim5}{W} for a \SI{15}{m}
stripe. When transmitting \SI{20}{Gbit/s} over an individual \gls{pmf} -- achievable with a 'gentle' low-constellation waveform in \SI{20}{GHz} bandwidth -- the energy efficiency of the \subthz transmission is estimated  to be \SI{\sim200}{pJ/bit}\footnote{Note that digital baseband power is not included in this estimate. We expect its contribution in energy/bit is rather agnostic to the operating frequency.}.
This shows that, in absolute terms, the energy required to provide multi-Gbit/s wireless data transmission is very reasonable. For example in a factory or a stadium, when compared to other consumers such as lighting and machinery, it would not incur a significant increase.} %

\subsection{Transmitting the \subthz signals over the air}

The \gls{pmf} deployment with many distributed \glspl{ru} \update{boosts the signals along the fiber} and brings the
antennas close to the \gls{ue}, thereby reducing the \gls{los} propagation distance and hence
the path loss \update{over the air}; \update{most importantly, }it creates redundant links to cope with blocking.  The spot-beam coverage also reduces
scan loss, since the smaller coverage means that the direction to the
\gls{ue} is closer to the broadside of the antenna.

To demonstrate this with an example \update{based on data from real \gls{pmf} prototypes
and \subthz RF component implementations~\cite{Strömbeck2023icd, 10258328}}, we consider a conference
room (\cref{fig:pl-example}), where a \SI{15}{\meter}-long stripe runs
along the length of the room, with each \gls{ru} providing
\SI{3}{\meter} of coverage along the horizontal ($x$)-axis. If the
\gls{ue} is located \SI{4}{\meter} below the stripe, with an antenna
gain of \SI{18}{dBi} (4 $\times$ 4 patch array with element gain of
\SI{6}{dBi} and array gain of \SI{12}{dB}), the path loss at different
positions is shown in \cref{fig:pl-example}. It is assumed that
handover occurs when the \gls{ue} is between two \glspl{ru}. The
\gls{ru} also has 4 $\times$ 4 patch array giving \SI{18}{dBi} antenna
gain. For comparison, the path loss of a system with only one
centralized \gls{ru} at the center of the room is also shown in
\cref{fig:pl-example}, for both an unsteered and steered case.

As can be seen, the increase in path gain (with respect to the
centralized \gls{ru}) begins when the handover occurs at
$x=\SI{1}{\meter}$, with the excess loss caused by both the increase
of over-the-air distance (path loss) and direction to broadside (scan
loss). The path loss of the distributed \glspl{ru} remains stable,
with regular handovers occurring along the $x$-axis in multiples of
\SI{1}{\meter}. In fact, near the edge of the room, the maximum path
loss from the central 4~$\times$~4 patch array is the same as in the
deployment with only single patch antennas along the stripe. This
means that the distributed \gls{ru} can be equipped with a single
patch element and still provide the same link budget as the
centralized system with a 4$\times$4 patch array at the \gls{ru}. %
\update{The single-element antenna approach implies a simpler, and often also more compact, design. %
 It does not require beam scanning and avoids the complexity and losses coming with phase shifting and feeding network. }The
unsteered stripe case could also be interpreted as a high-gain single
antenna~\cite{10258328}.

Finally, the distributed \glspl{ru} also offer larger
probability of \gls{los} propagation as  the closest \gls{ru} is
located at a higher elevation angle with respect to the \gls{ue}, compared
to the centralized case. The overlapping coverage further increases the probability of \gls{los} propagation.

\textbf{Cost considerations.} \update{The envisioned ease of installation of the \gls{pmf} is a major benefit. Ultimately, the fibers could be deployed by simply gluing them as stripes of tape that embed the (very small) \glspl{ru}. Installation costs can be safely assumed to be the dominant cost of any wireless infrastructure with a distributed topology. The latter is the only viable option for providing consistent coverage at \subthz, as explained above. An alternative could be to deploy optical fibers, which present a low-loss solution yet would require many expensive optical-electrical converters. Bringing the digital baseband signals to the distributed entities would require many costly full \subthz front-ends; as direct electrical interconnects cannot support the  bandwidth required for multi-\unit{\giga bit/s} datarates, this is not a viable alternative. Note that no quantitative cost estimates are provided as these depend highly on market demand and volume. }

\section{Strengthening \subthz links Concept II -- Dual-frequency for reliability, \update{efficiency}, and capacity}\label{sec:dual}

Fundamental properties of lower and higher bands in a dual-band system
are complementary. Lower bands offer more modest
throughput -- due to comparatively lower bandwidth -- but higher link
robustness, that is, less sensitivity to blockage, \update{and Doppler}. Operation at higher
bands has precisely opposite qualities: high throughput but higher
fragility of the links.  Hence, it is natural for lower and higher
bands to take on complementary roles in a dual-band system.

Control data typically requires high link robustness and should be transmitted on the \subghz band:
scheduling, configuration of transmission and reception, \update{handovers between \glspl{ru}, wake-up functionality for stripes temporarily put in sleep mode},
etc. Lower bands can also be used to transmit data that
require high reliability. On the other hand, the connection at the
\subthz band should almost exclusively be used for data
transmissions. Non-data signaling at higher bands should be limited to signals that cannot be sent
at the lower band, e.g., transmission of pilots for estimation
and tracking of RF impairments.  Moving the control signaling to the lower band not only ensures the necessary robustness but  may also reduce the control
signaling overhead and latency, e.g., through the selection of the best \gls{ru} and associated beams at \subthz learned at \subghz.

To understand the underlying problem, note that
since higher bands typically use phased antenna arrays with few
transceiver chains, beamforming in several directions simultaneously
may be infeasible. Instead, focused beams can be directed only in a
single direction at a given time. Finding the best \gls{ru} and beam
to serve the \gls{ue} then entails sending one pilot signal per \gls{ru}
and candidate beam in a separate time slot, and the \gls{ue} performing
measurements on each pilot signal and providing a measurement
report. Assuming $N_{RU}$ candidate \glspl{ru} and $N_{b}$ candidate
beams per \gls{ru}, the described procedure requires $N_{RU}N_{b}$ time
slots for sending the downlink pilots, plus \gls{ue} processing and
reporting time. If the number of \glspl{ru} and beams per \gls{ru} is large %
the
described procedure could end up consuming a substantial amount of
downlink resources and result in large latency.

Lower bands can be conveniently used for improving the efficiency of
the above procedure. The solution entails two stages, a learning
phase, which can be performed as for example proposed
in~\cite{9557817}, and an exploitation phase:
\begin{enumerate}
\item During the learning phase, a legacy measurement and reporting
  scheme is performed at the \subthz band. Simultaneously, the dual-band UE sends uplink pilots at \subghz band
  to estimate the channel. The \subghz channel and
  information about the best \subthz \gls{ru}/beam combination provide a
  data point particular to where the dual-band \gls{ue} is
  located. This data point is used by the network to learn the mapping
  from the channel at \subghz to the choice of the best \gls{ru}/beam at
  \subthz. Collection of data points in the
  learning phase continues over time and \gls{ue} positions until
  satisfactory performance of the model is achieved.

\item In the exploitation phase, when there is need to choose a new \gls{ru}
  or beam, the \gls{ue} first sends an uplink pilot in the \subghz band. Using
  the model inferred in the learning phase, the network infers the
  list of \subthz \glspl{ru}/beams most likely to be the best for serving the
  \gls{ue}. This list can be utilized for a more concentrated \gls{ru}/beam
  search, or the best high frequency \gls{ru}/beam as
  suggested by the model could be used directly for data transmissions
  without measurements.
\end{enumerate}

\update{Supervised methods using standard machine learning solutions
  (e.g., deep feedforward or convolutional neural networks) can be
  used to learn the mapping from \subghz channels to the best \subthz \glspl{ru} and beams~\cite{9121328}. During training, the \subghz channel responses represent  features. The  best \subthz \glspl{ru} along with the best beam represent a label in a feature-label pair. The size of the dataset for training is determined by the coverage area, wavelength, and correlation properties of the \subghz channel. For example, sampling a coverage area of $50\times 50$ meters on a regular 2D grid with \SI{5}{\lambda} spacing at \SI{6}{GHz} yields \num{40000} samples. Preliminary results~\cite{NishantGuptaVTC} suggest that a significant reduction in overhead for finding the best \subthz \glspl{ru} can be achieved with this approach.}

 \update{One challenge with the dual-band operation is the need for tight synchronization and coordination between the two bands. This can be accomplished  by having the \subghz and \subthz systems share the same digital baseband and scheduler. Moreover, the time-domain structure of the two subsystems should be designed such that joint scheduling is facilitated. For example, each timeslot in the \subghz system could contain an integer number of \subthz slots.} %

   \section{Conclusions and R\&D directions} %

\Subthz systems will not offer robust services to non-static
terminals when deployed similarly as \subghz
systems.
We have presented
a novel \update{dense RF-over-\gls{pmf}-based distribution of \subthz signals} and dual-frequency concept for offering robust \subthz connectivity. This system \update{provides a low-cost solution for a distributed deployment and}
circumvents propagation and
hardware obstacles
encountered at \subthz frequencies. The proposed approach offers an interesting
potential to strengthen \subthz links. \update{The system can extend coverage to non-static \glspl{ue} over tens of meters with an energy efficiency in the order of a few \SI{100}{pJ/bit}. } %

Further technological \update{innovation is} needed to advance the concept
towards actual implementation:
\begin{itemize}

    \item Lower loss fibres and couplers should be designed to improve the link budget over the \gls{pmf}. \update{This would require less amplification of \glspl{ru} in booster mode, lowering energy consumption.}

    \item Efficient algorithms are needed to enable transmission of
      waveforms with non-linear and noisy hardware. \update{Appropriate waveforms are required to relax both hardware and algorithmic complexity.}

     \item Approaches
       leveraging  the same dual-frequency dense deployment concepts
       should be developed to support emerging applications that require position information.
\end{itemize}

After decades of R\&D on wireless transmission at \mmwave
and above, these systems have so far only found success on \update{fixed or very short links}.
Other trajectories are needed to conquer new territories.  While significant
technological progress is needed to prepare our proposed dual-frequency
approach for actual deployment,
\update{we think that this is the way forward to exploit the large bandwidths at \subthz bands.}

\section*{Acknowledgments}
{\footnotesize
The authors would like to thank the 6GTandem consortium
for discussions of the system concept and hardware R\&D.
}

\printbibliography

\clearpage
\begin{IEEEbiographynophoto}{Liesbet Van der Perre}[SM]\,(liesbet.vanderperre@kuleuven.be) is Professor at the department of Electrical Engineering at the KU Leuven in Belgium and a guest Professor at the university of Lund, Sweden. She was with the nano-electronics research institute imec in Belgium from 1997 till 2015.
Her main research interests are in wireless communication and embedded connected systems. She is involved in several projects progressing indoor positioning, IoT, and 6G and beyond 6G for sustainable applications and takes up the role as scientific coordinator of R\&D partnerships.
\end{IEEEbiographynophoto}

\begin{IEEEbiographynophoto}{Gilles Callebaut}[M]\,(gilles.callebaut@kuleuven.be) earned his M.Sc. degree in Engineering Technology from KU Leuven, Belgium, in 2016, graduating summa cum laude. He completed his Ph.D. in Engineering Technology at the same institution in 2021, focusing on optimizing energy efficiency in \gls{iot} devices through single and multi-antenna technologies. He is currently an assistant Professor at the Electrical Engineering and Architecture Department, KU Leuven, Belgium. He focuses on creating scalable, sustainable, and energy-efficient wireless technologies for \gls{iot}, 6G, and beyond, while actively promoting transparency and collaboration through open science. His main focus areas are: sustainability, \gls{iot}, 6G, WPT over \gls{rf} with multi-antenna systems and energy-efficient design.
\end{IEEEbiographynophoto}

\begin{IEEEbiographynophoto}{Thomas Eriksson}[M]\,(thomase@chalmers.se)  
received his Ph.D. in information theory from Chalmers University of Technology, Sweden, in 1996. He worked at AT\&T Labs-Research, USA (1997–1998) and Ericsson Radio Systems, Sweden (1998–1999) before joining Chalmers in 1999, where he is now Professor of communication systems and Vice Head of the Department of Electrical Engineering. He was Guest Professor at Yonsei University, South Korea (2003–2004). He has authored over 300 papers, holds 14 patents, and leads research on hardware-constrained communications, focusing on massive MIMO, hardware impairment compensation, data compression, and DPD for efficient wideband transmitters.
\end{IEEEbiographynophoto}

\begin{IEEEbiographynophoto}{Muris Sarajlic} (muris.sarajlic@ericsson.com)
was born in Tuzla, Bosnia and Herzegovina, in 1987. He received the B.Eng. degree from the Department of Electrical Engineering, University of Tuzla, Bosnia and Herzegovina, in 2010, and the M.Sc. and Ph.D. degrees from the Department of Electrical and Information Technology, Lund University, Sweden, in 2013 and 2019, respectively. In 2019, he joined Ericsson Research. His research interests include energy efficiency and the complexity aspects of hardware-constrained wireless communication systems.
\end{IEEEbiographynophoto}

\begin{IEEEbiographynophoto}{Christian Fager}[F]\,(christian.fager@chalmers.se) received his Ph.D. degree from Chalmers University of Technology, Sweden, in 2003 where he is Full Professor since 2019. He has authored/co-authored more than 250 publications, making significant contributions to the areas of power amplifiers and wireless communication transmitters. He served as the 2021 IEEE PAWR Conference Chair and as Associate Editor for IEEE Microwave Magazine and IEEE Microwave and Wireless Technology Letters. He is Member of the Board of Directors of the European Microwave Association (EuMA) and the IEEE MTT-S Technical Committee on Wireless Communications. Dr. Fager received the Chalmers Supervisor of the Year Award in 2018.
\end{IEEEbiographynophoto}

\begin{IEEEbiographynophoto}{Fredrik Tufvesson}[F]\,(fredrik.tufvesson@eit.lth.se)
received his Ph.D. in 2000 from Lund University in Sweden. After two years at a startup company, he joined the department of Electrical and Information Technology at Lund University, where he is now professor of radio systems. His main research interest is the interplay between the radio channel and the rest of the communication system with various applications in 5G/6G systems such as massive MIMO, distributed MIMO, mm wave communication, vehicular communication, integrated communication and sensing, and radio-based positioning.
\end{IEEEbiographynophoto}

\begin{IEEEbiographynophoto}{Buon Kiong Lau}[SM]\,(buon\_kiong.lau@eit.lth.se)
is a Professor and Head of the Communications Engineering Division at Lund University. He is most recognized for his contributions to multi-antenna systems, on aspects relating to antennas, propagation, and signal processing. He has held several editorial roles with IEEE journals and various committee roles in the IEEE Antennas and Propagation Society. His career spans research, teaching, and leadership, with over two decades of experience across academia and industry. His work bridges theory and practical deployment, with a strong focus on advancing wireless communication technologies through interdisciplinary research.
\end{IEEEbiographynophoto}

\begin{IEEEbiographynophoto}{Erik G. Larsson}[F]\,(erik.g.larsson@liu.se)
is Professor at Link\"oping University, Sweden, and a Fellow of the IEEE and EURASIP.  He co-authored \emph{Fundamentals of Massive MIMO} (Cambridge, 2016) and \emph{Space-Time Block Coding for Wireless Communications} (Cambridge, 2003). He received the {IEEE Signal Processing Magazine} Best Column Award twice, in 2012 and 2014, the 2015 IEEE ComSoc Stephen O. Rice Prize in Communications Theory, the 2017 IEEE ComSoc Leonard G. Abraham Prize, the 2018 IEEE ComSoc Best Tutorial Paper Award, the 2019 IEEE ComSoc Fred W. Ellersick Prize, and the 2023 IEEE SPS Donald G. Fink Overview Paper Award.
\end{IEEEbiographynophoto}

\end{document}